# Building a Digital Twin for British Cities

Michael Batty and Richard Milton

Centre for Advanced Spatial Analysis,
University College London, 90 Tottenham Court Road,
London W1T 4TJ, UK

m.batty@ucl.ac.uk  richard.milton@ucl.ac.uk
 @jmichaelbatty

3 December 2023

## Abstract

Ever faster computers are enabling us to extend our standard land use transportation interaction (LUTI) models to systems of cities within which individual cities compete for resources within the wider environment in which they interact. As we scale up in this way, we are able to simulate and measure the impacts of large-scale infrastructures at different spatial levels. Here we build a platform, which is essentially a digital twin, for over 8000 urban places in Great Britain where we can rapidly model all flows between these locations using multi-modal spatial interaction models. We first present the structure of the model and then apply it to population, employment and trip flow data for three modes of travel (road, bus and rail) between small spatial units defining the three countries, England, Scotland and Wales. We then tune and train the model to reproduce a baseline, and follow this with a demonstration of the web-based interface used to run and interact with the model and its predictions. Once we have developed the platform, we are able to explore variants of the twin, partitioning the country in different ways, showing how different forms of spatial representation change the performance of the model. We are developing the model at a much finer scale making comparisons of performance while adding another mode – an active travel layer – that elaborates the twin. We finally illustrate how the model can be used to measure the impacts of new scenarios for rail, simulating the Integrated Rail Plan and the High Speed 2 (HS2) proposal.

*Acknowledgements:*
The project has been funded by The Future Cities Catapult (2016-2018),
and the Alan Turing Institute under Contract–CID–3815811.

# Introduction

A digital twin is traditionally defined as a computer model of a physical system that closely interacts with the real system in question. This enables the operation of the real system to be sensed by the twin and for the digital model to control and monitor feedback that ensures the real system's operation meets certain objectives that sometimes imply optimal performance. The twin usually provides some predictive capabilities with respect to the dynamics of the real system and is sometimes used in the development of new designs for the physical system itself. The dominant type of twin in this definition is a digital version of an electro-mechanical system, a physical machine for example such as an aircraft or an automobile, although the great flurry of interest in the idea of digital twins in the last decade has broadened this definition considerably. Many socio-economic systems built around ideas, methods, and policies, which are traditionally non-digital, are beginning to acquire the status of digital twins. Computer models of such systems, however, go back to the dawn of electronic computing itself but only in the last decade have digital models of very large scale social systems begun to proliferate.

What has also emerged are multiple digital models of the same system, each purporting to represent some features of the real system, but differing in their emphasis and thus forming collectives of twins that together define different methods of control, prediction, and design. The idea that there can only be one twin is now passé for it is increasingly clear that as models are simplifications of the real thing, there can be many different variants mirroring different features of the real system that embody different theoretical perspectives. There are many elements of a single model that can be changed in various ways through its representation of the real system in question or through changes to its parameters or both, and this leads to different twins being variants of the same model. As new models are spawned in this way, we generate a 'cornucopia' or 'federation' of twins that need to be coordinated with respect to problems and challenges that face the real system.

The notion that there should always be more than one model of the same system was first articulated more than 50 years ago. Several variants of econometric model were built for national economies each giving different weight to certain factors and generating a range of forecasts. Many national banks and treasuries quite routinely now use a range of models for prediction that generate an ensemble of forecasts that are pooled to produce more accurate outcomes (Bates and Granger, 1969; Wikipedia, 2023a) while in weather forecasting, the same sorts of ensemble approach are widely used to explore short and medium term forecasts (Blum, 2019; Thompson, 2022). These are essentially sets of digital twins that all address the same problem but combine and emphasise different approaches and data associated with different variants of forecasting. It is an acknowledgement that there is no one approach that is preferable to any other (Wu and Levinson, 2021) and all are needed to generate a requisite understanding for modelling complex systems. In 1976, Greenberger, Crenson and Crissey argued that more than one model of the same system always needs to be built so they can be pitted against one another – 'counter modelling' they called the process - while 40 years later, Page (2019) argued for a 'many-model' paradigm to underpin all our approaches to prediction in ill-defined systems.

Computer models of cities are excellent examples of this broader definition of digital twins. There are a plethora of such models, ranging from simple digital representations of the 3-dimensional form of cities, that often feature in problems where form dominates function, all the way to models of how local urban economies can be conceived of as spatial flows. These latter models are associated with rent, income, revenues, prices and so on that underpin how cities function as sets of economic markets and how inequalities manifest themselves in the distribution of various assets, housing for example. Models that can be represented spatially can usually be aggregated to different scales, and at each level of aggregation, the nature of the representation can change while the functioning of the model remains the same. In our discussion of digital twins here, we will focus very strongly on scale and how changes in representation generate different twins but which remain rooted to their parent.

One of the key features of the digital twin we will build for all cities in Great Britain is that we are able to deal with a relatively closed system, geographically that is. In a global world which is rapidly urbanizing across national boundaries, defining a system relative to its environment where the interactions between each are at a minimum is difficult. However by scaling up from individual cities to systems of cities as Berry (1964) defined the spatial world over 60 years ago, we are able to work with relatively closed systems where interactions both within and between cities are central to the model. Moreover by the end of this century, the global population will be largely urban with over 95% living in cities of one size or another. The digital twin of British cities that we report in this paper is one of the first to grapple with how we can scale our traditional city models to much larger entities where many cities grow, fuse and evolve together: our twin for urban Britain thus represents a prototype for dealing with the core elements of what Brenner (2013) amongst others refers to as 'models of planetary globalisation'.

In this paper, we will first present the structure of the urban model which we call QUANT[1] that has been built for Great Britain (Batty and Milton, 2021). It is essentially an equilibrium structure in which employment is a function of population and population a function of employment. We break into this simultaneity from employment which is the driver of population, in turn constrained by locational capacities and regulations on development. We then illustrate how the model converges to an equilibrium where the linkages are modelled using spatial interaction models of the gravitational type. We can extend this model structure into several other sectors such as retailing, education, and health but in the applications we develop here, we focus entirely on residential location.

The generic Land Use Transportation Interaction (LUTI) model will be presented in the first main section. This is a structure where population and employment in Great Britain are located in small zones which in the standard Census geography are called 'output areas'. This first baseline twin is thus composed of zones at intermediate scale called Middle-layer Super Output Areas (MSOAs). Interactions between these zones represent the glue used to stitch these locations together using gravitational principles which embody widely applicable inverse distance relationships. For our baseline twin, there are some 8436 MSOAs which generate flow or interaction matrices of the order of 71 million trips and this gives an idea of the scale of the application. The rest of the paper deals with variants of this digital twin, and in the second section, we show the difference between the spatial data in the three countries of England, Scotland and Wales.

In the third section, we train or tune the model to GB by ensuring that the average travel times for each mode of travel reproduce those that we observe from the data. This results of this tuning are then compared against the locational and interaction data as a series of quasi-independent tests of how well the model performs. In the fourth section, we show the way we interact with the model through the web-based interface that begins with model exploration, moves to calibration and testing, then to providing a basis for using the model predictively, primarily in generating and testing 'what-if?' scenarios. In the rest of the paper, we explore variants of the twin, in terms of changing its spatial representation, first by partitioning the GB model into different scales and then showing how representation changes using different scenarios. In the fifth section, we illustrate the generic structure for Greater London, first at MSOA level and then at a finer level which is five times smaller for zones called Lower Super Output Areas (LSOAs). The change is significant given the model is the same one but at different scales and the performance of the London subset differs by some 7 percent from the complete GB model. We then add a new transport layer to the generic model introducing a 'walk-cycle' mode into the Greater London and this changes the performance once again in significant but not radical terms.

---

[1] **Q**uantitative **U**rban **AN**aly**T**ics. Great Britain (GB) is defined as England, Scotland, and Wales.

In the sixth and penultimate section, we further change our model representation developing another variant of the twin exploring the impact of the 2022 Integrated Rail Plan on locations and interactions between all zones. The Plan consists of the new High Speed 2 line from London to the Leeds via Birmingham and Manchester and the impact of trimming this line to first exclude Leeds, then Manchester, and finally by ending the line in west London. At the time of writing, this proposal is a political hot potato and besides showing how changes in the distribution of these resources leads to very different savings in travel time, the changes that result from these applications provide an instructive analysis of how we can build many variants, many twins, of the same system. We conclude with a brief evaluation of the problems involved in extending our twin to an entire country, and speculating on how such models might evolve as data gets better and computation ever faster.

## The Structure of the Urban Model

The model defines the urban system in terms of spatial interactions between different activity sectors whose elements are represented in small zones. These are locations of employment $E_i$ and population $P_j$ in zones $i$ and $j$ but there are several variants of this model that include additional activities such as education, health and retailing, simple extensions of the model which link population and employment to these activities in terms of other types of spatial interaction (Spooner et al., 2021; Lopane et al., 2023). The structure of the model defines two sets of processes. First there are those that link activities in terms of functional relations which we specify generically as $E = g(P)$ and $P = h(E)$ and which define the circularity of the system as a set of simultaneous equations $E = g(h(E))$ and $P = h(g(P))$. Second these activities depend on each other spatially in that population is generated by employment and employment by population through spatial interactions which are modelled using gravitational or discrete choice models. Here we will use gravitational models where we break employment and population into flows defined here as probabilities of interaction $p_{ij}$ between zones $i$ and $j$. In their generic form, these can be represented as $p_{ij} \propto E_i P_j f(c_{ij})$ where $(c_{ij})$ is a function of the travel cost (or time or distance travelled) between employment and population locations.

The full model can be stated as a concatenation of probabilities of interaction where the two sectors are linked as follows: for the employment sector

$$T_{ij} = E_i p_{ij} \qquad (1)$$

where $\sum_j T_{ij} = E_i$ and the conditional probability of working in $i$ and living in $j$ is normalised to $\sum_j p_{ij} = 1$. $T_{ij}$ is the flow or trips from place of employment $i$ to residential location $j$ and the population $P_j$ attracted to $j$ is then computed as

$$\sum_i T_{ij} = P_j . \qquad (2)$$

The population in $j$ links to another sector, typically back to the employment in zones $k$, as

$$S_{jk} = P_j q_{jk} \qquad (3)$$

where $\sum_k S_{jk} = P_j$ and the conditional probabilities of living in $j$ and engaging in $k$ are normalised to $\sum_k q_{jk} = 1$. $S_{jk}$ is the trip demand from place of residence $j$ to employment location $k$ and the employment demand $E_k$ attracted to $k$ is computed as

$$\sum_j S_{jk} = E_k . \qquad (4)$$

In this model, the flow from residential locations back to employment centres is not necessarily the reverse journey to work but the demand by the population for goods and commodities available in retail centres. If we equate $E_k$ with $E_i$, then we invoke the simultaneous nature of the employment-population circularity but it is preferable to articulate this simply as an additional sector that closes on employment but which is not the same as the employment which drives the cycle beginning in equation (1).

It is worth noting that to solve the model in equations (1) to (4) in simultaneous fashion, we ensure convergence by solving these equations iteratively using the following sequence:

$$\left. \begin{array}{l} P_j(t) = \sum_i T_{ij}(t) = \sum_i E_i(t) p_{ij} \\ E_i(t+1) = \sum_j S_{ji}(t) = \sum_j P_j(t) q_{ji} \end{array} \right\} . \qquad (5)$$

Once the cycle from time or iteration $t$ to $t+1$ is complete, we substitute $E_i(t+1)$ for $E_i(t)$ in equation (5) and we reiterate until convergence. This convergence is not proven but the equations for this model are sufficiently robust that in all the examples we (and others) have experimented with so far, convergence, hence the simultaneity of the system, has been assured. Lowry (1964) illustrated this type of convergence in one of the first models developed in this tradition for the city of Pittsburgh in the early 1960s.

The variant of the model that is developed here only uses the employment-population sector which we refer to as the residential location model. Using the conditional probability form, we can elaborate the model in two ways. First, we use the model to represent different transport modes that we define by the index $m$, and second, we introduce constraints on the predicted population $P_j$. The attraction of each residential zone is a measure of size such as floorspace $A_j$ while $c_{ij}^m$ is the travel time from $i$ to $j$ on mode $m$ and $\lambda^m$ is the relevant parameter. We can state the basic residential location model disaggregated by different transport modes as

$$T_{ij}^m = E_i p_{ij}^m = E_i \frac{A_j \exp(-\lambda^m c_{ij}^m)}{\sum_z \sum_j A_j \exp(-\lambda^z c_{ij}^z)} \qquad (6)$$

where it is clear that total trips over all modes are

$$T_{ij} = \sum_m T_{ij}^m = E_i p_{ij} = E_i \sum_m p_{ij}^m = E_i \frac{\sum_m A_j \exp(-\lambda^m c_{ij}^m)}{\sum_m \sum_j A_j \exp(-\lambda^m c_{ij}^m)}, \quad \sum_m p_{ij}^m = p_{ij}, \tag{7}$$

and the modal split on each link $i \leftrightarrow j$ is

$$\frac{T_{ij}^m}{T_{ij}^z} = \frac{\exp(-\lambda^m c_{ij}^m)}{\exp(-\lambda^z c_{ij}^z)}. \tag{8}$$

The second variant of the residential location model is semi-constrained to meet given constraints on the capacity $Z_j$ of each zone to receive population. The spatial interaction equation can thus be defined as

$$T_{ij} = \sum_m T_{ij}^m = \frac{\sum_z B_j A_j \exp(-\lambda^z c_{ij}^z)}{\sum_z \sum_j B_j A_j \exp(-\lambda^z c_{ij}^z)} \tag{9}$$

where $B_j$ is the weight imposed on the residential location $j$ to ensure that any constraint on capacity given by $Z_j$ is met. If $B_j = 1$, then there is no constraint on the trips attracted to zone $j$ where $P_j = \sum_i T_{ij}$, otherwise $B_j$ is chosen to ensure the inequality is met, that is $P_j \leq Z_j$. The iterative procedure for ensuring that these inequality constraints are met is defined in Batty and Milton (2021).

Representing the Twin in Geospatial Data

The digital twin for England, Scotland and Wales (GB) is a geospatial model of demographic and economic activity[2] whose representation is based on some $n = 8436$ census tracts defined earlier as Middle-layer Super Output Areas (MSOAs). The average employment in these areas is 2378, the population 7270 while the average area of each zone is 27.73 square kms (or 2773 hectares) giving an overall population density of 2.61 persons per hectare. With GB divided into these zones, the magnitude of flows between them is of the order of $n^2 (= 71,166,096)$ and in the basic twin which is our starting point, there are three sets of such flows related to road, bus and rail. The total employment in the country in 2011 was 20,060,434, the population 61,330,712 and thus there are more potential flows between all areas than total population. When the number of zones increases by almost 5-fold to the next finest census layer, the Lower-level Super Output Areas of which there 41,729 zones, the total number of potential flows is 1,741,309,441 (1.7 billion). The model we will present here is based on the coarser MSOA level but we have developed a variant of the model – another twin – at this finer level of granularity. It will be discussed in a forthcoming paper.

These models have been applied mainly in cities with more than 1 million population in the last 50 years but only recently have computational resources become available to develop them for very large numbers of zones, sufficient to represent the level of detail necessary for appropriate strategic forecasting and planning. The digital twin we describe here began as a model for Greater London and was then extended to London and its outer metropolitan area, scaling from some 7 to 15 million population (Batty, 2013a; Batty et al., 2013). However,

---

[2] These are models are referred to generically as Land Use Transportation Interaction (LUTI) Models although they mainly deal with economic and demographic activity associated with land use and transportation, see https://assets.publishing.service.gov.uk/government/uploads/system/uploads/attachment_data/file/939135/tag-supplementary-luti-models.pdf

Britain is a highly centralized urban system and many of its largest cities – the so-called core cities – are intimately related to London; in this sense, as we scale the model, we take in all the significant urban places in the country (Haldane and Rees, 2023). To an extent, applications to systems of cities which comprise individual cities are in the forefront of change (Berry, 1964) while the system we are twinning with here, is a relatively closed, somewhat compact region that can be treated as a single entity. The fact that it is an island provides a measure of closure that enables it to be treated as a whole. The MSOA zoning system is shown in Figure 1(a) where it is clear that the urban areas contain much more compact and smaller zones than the rural. Although this figure is simply the set of polygons describing how the country is subdivided into zones, the clustering of these zones provides a good impression of the way cities are joined together as well as the way they spread out. Figure 1(a) thus resembles the map of urban development.

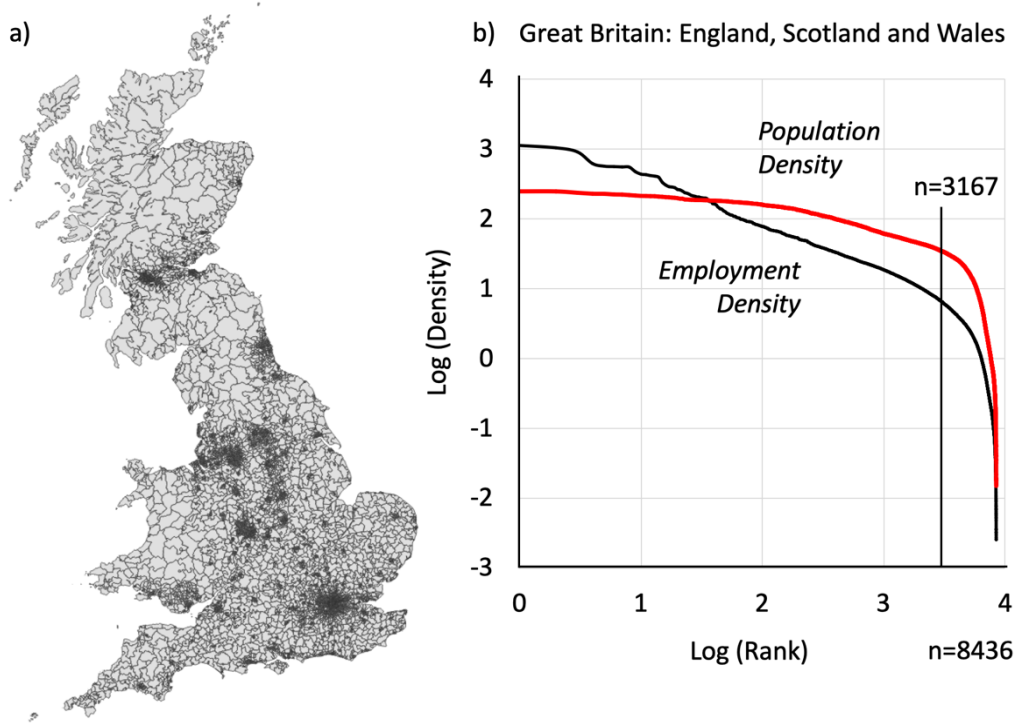

*Figure 1: The MSOA Subdivisions of GB and Log Rank-Size of Population and Employment Densities*

The distribution of employment and population activities in the zones shown in Figure 1(a) follow inverse relationships which, when plotted as logarithmic rank-size relations, imply distributions close to power laws especially in their upper tails. We show this for the whole of Great Britain in Figure 1(b) and this implies a distribution closer to the log-normal than the inverse power; but if we truncate the distribution at the point where the line appears to straighten out, then the relationships follow an almost a perfect inverse power law. We first fit this relation to the employment densities (for the sum over all modes of travel) for the top 3167 cities/zones as $[E_i/L_i](r_i) = K r_i^{-\alpha}$ where $r_i$ is the rank of employment density in zone $i$, $K$ is a constant of proportionality close to its theoretical value $K = [E_1/L_1](r = 1)$, $L_i$ is the land area of zone $i$, and $\alpha$ is a power. This equation is estimated in log-linear form as $\log[E_i/L_i](r_i) = \log K - \alpha \log r_i$ where the parameters of the distribution are estimated as $\log K = 3.390$ and $\alpha = 0.721$ and where the r-squared value (the proportion of variance explained) is 0.988. We can also fit the population density distribution using the same logic and this yields the log-linear equation $\log[P_i/L_i](r_i) = \log G - \beta \log r_i$. Here the parameters of the distribution are $\log G = 3.034$ and $\beta = 0.419$ and the r-squared value (the proportion of variance explained) is 0.953. It is clear from these statistics and from Figure 1(b) that the employment density is much more concentrated (steeper in terms of the graph) than the population and this implies that the biggest zones in terms of employment are in the smaller number of larger city centres.

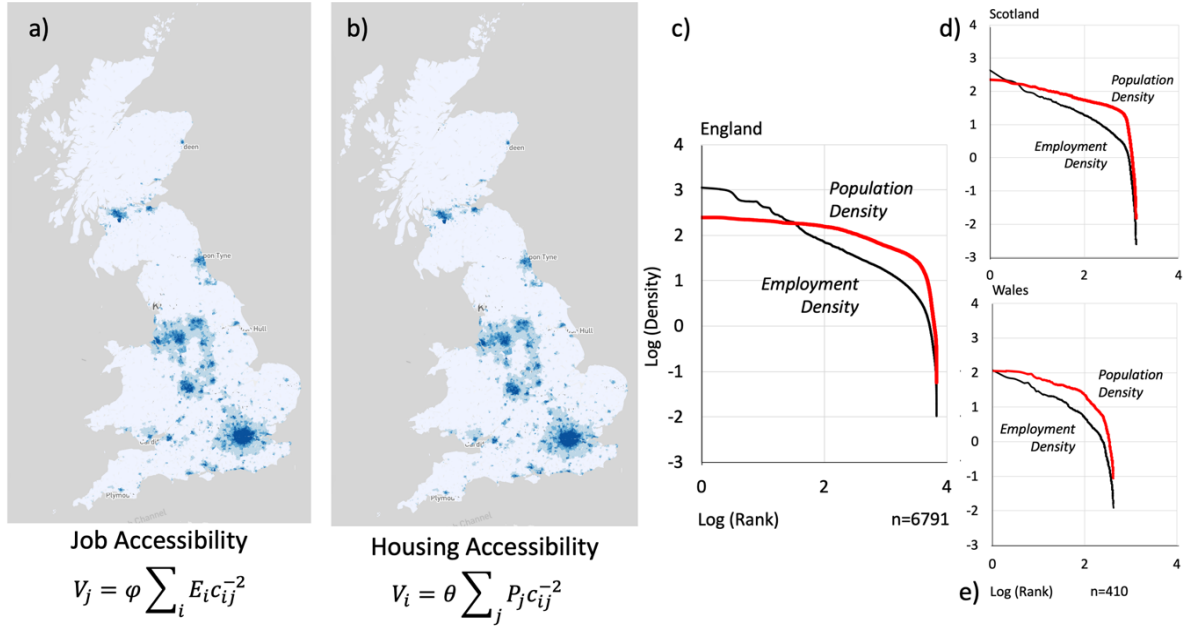

Figure 1: Job and Housing Accessibility Densities and Log Rank-Size
of Population and Employment Densities for England, Scotland and Wales

There are quite significant differences between the three nations that compose GB. Figure 2 shows the distribution of employment and population in the three nations but smoothed and aggregated so that the relative skewness is dampened by computing accessibilities to jobs $V_j$ and residences $V_i$. These accessibilities are defined as $V_j = \varphi \sum_i E_i c_{ij}^{-2}$ and $V_i = \theta \sum_j P_j c_{ij}^{-2}$ where $\varphi$ and $\theta$ are constants that enable the distributions to be plotted in suitably normalised fashion. These two accessibility distributions are almost identical and only if the reader zooms-in on these maps can differences be discerned. From these visualisations, it is clear that the overall density of employment in Scotland is less than Wales which in turn is less that England. In terms of population, England has some 88% of the total, Scotland near 8% and Wales some 4%. The Scottish urban system appears somewhat different in the distribution of employment densities from England and Wales and this is probably due to climate, natural resources, cultural factors and laws. The employment densities also reflect key differences with respect to the development of the three nations and although all three indicate strong inverse relationships between density and rank, the urban system is much more concentrated in England than in Wales, and then Scotland. The easiest way to compare these differences is presented in Table 1 where we show all the regression estimates for GB and the three nations that form the whole country.

The relative proportions of employment and population in the three nations are reflected in the densities of the zones. Of the top 100 MSOAs in GB, in terms of employment density, some 89 are in England, 9 in Scotland and 2 in Wales. The correlation between employment and population densities measured by the r-square statistic is very low at 0.068 and this suggests that any model linking population and employment should be articulated using explicit interaction functions of the spatial variety such as those we have already stated as being based on gravitational hypotheses. In short, we cannot explain employment or population density in simple linear terms linking one another, so simulating the interactions between them is essential. Thus in our full model, we introduce other sets of data based on interactions between MSOAs by different modes. To an extent of course, this focus on interaction has always been the rationale for explaining location in terms of modal accessibilities (Batty, 2013b).

*Table 1: Regression Coefficients for Truncated Density Distributions for GB and Partitions*

|  | $GB^+$ $n = 8436\ (3167)$ | | $England$ $n = 6791\ (2980)$ | | $Scotland$ $n = 1235\ (720)$ | | $Wales$ $n = 410\ (130)$ | |
|---|---|---|---|---|---|---|---|---|
|  | Emp Density | Pop Density | Emp Density | Pop Density | Emp Density | Pop Density | Emp Density | Pop Density |
| Intercept $\log K$ | 3.390 | 3.034 | 3.208 | 2.824 | 2.659 | 2.450 | 2.197 | 2.230 |
| Parameter $\alpha$ | -0.721 | -0.419 | -0.663 | -0.338 | -0.699 | -0.349 | -0.711 | -0.400 |
| r-square $R^2$ | 0.988 | 0.953 | 0.996 | 0.898 | 0.986 | 0.995 | 0.977 | 0.967 |

+ The number of MSOAs used to fit the power laws are given in (brackets) after the number of MSOAs

The data associated with the flows (called interactions or trips) from zones of employment to residential locations are available from the UK Population Census where we have used the latest available version from 2011. In fact we have scaled all our data to 2021 values from intermediate estimates which form the baseline date for the twin we are building here. The flows are from the Travel to Work Tables of the Census which are disaggregated by mode where we use road, rail and bus for this first twin, and walk and cycle (which we call active travel) for a later version of the model. The flow data needs to be allocated to the underlying networks which we take from sources such as Open Street Map, Ordnance Survey (OS) Mastermap and various *ad hoc* flow data from local Transportation Authorities while the travel times are taken from OS over the road measurements, and the GTFS (General Transit Feed Specification) for public transport and rail estimates. The flow data generates matrices of the order of 71 million links as we noted above but these are underpinned by networks between MSOA centroids which are fixed to the underlying geography which are at a much finer scale than the links between the centroids. The networks at the finest scale consist of 8.4 million segments and 3.5 million nodes for road, 0.420 million segments and 0.29 million nodes for bus, and 10,269 segments and 3,165 nodes for rail. The active travel layer is at a finer grain than the road network with 10.039 million segments and 8.25 million nodes. The centroids take account of the internal geometry of each MSOA while the intrazonal travel times $c_{ii}^k$ for each mode $k$ associated with centroid $i$ are computed from

$$c_{ii}^k = \frac{1}{s^k}\sqrt{\frac{L_i}{\pi}} \quad , \tag{10}$$

where $s^k$ is the average speed on mode $k$ and $L_i$ is the land area associated with zone $i$. Note that we measure the land area as that which is developed within the MSOA and in the case of disconnected patches of development, these are used to weight the travel costs.

The activities – employment and population – are stitched together in the model through the three modal networks $c_{ij}^k$ which compete for trips $T_{ij}^k$. These respond to changes in travel costs which in turn depend on the geometry and capacity of the underlying networks. These networks and interaction patterns differ considerably in the morphology and size of their flows and we can visualize these in different ways. It is not possible to visualize the thousands of flows for each mode but we are able to approximate these flows for each origin or destination as averages. For any flow $T_{ij}^k$, the displacement in space is defined by the coordinates from $i$ to $j$, $[x_i, y_i]$ to $[x_j, y_j]$ and their differences with respect to origin $i$ as $\Delta x_i = [x_i - x_j]$ and $\Delta y_i = [y_i - y_j]$ ) or equivalently to their destination. We will work with origins here which are employment zones and we first define these differences with respect to the length (magnitude) of their flows as $\delta_{ij} = [(x_i - x_j)^2 + (y_i - y_j)^2]^{1/2}$ where we take the differences in normalised coordinate form as $\Delta x_i' = [x_i - x_j]/\delta_{ij}$ and $\Delta y_i' = [y_i - y_j]/\delta_{ij}$. This transformation gives the orientation of each flow $T_{ij}^k, j = 1, 2, \ldots, n$ and we then weight the coordinate of each flow as $\overline{\Delta x_i} = \sum_j T_{ij}^k \Delta x_j'$ and $\overline{\Delta y_i} = \sum_j T_{ij}^k \Delta y_j'$. The average flow from each origin is given by the coordinates $x_i$ to $x_i + \overline{\Delta x_i}$ and $y_i$ to $y_i + \overline{\Delta y_i}$, but to visualize these patterns as below, we need to scale the entire system to reflect the geography of GB.

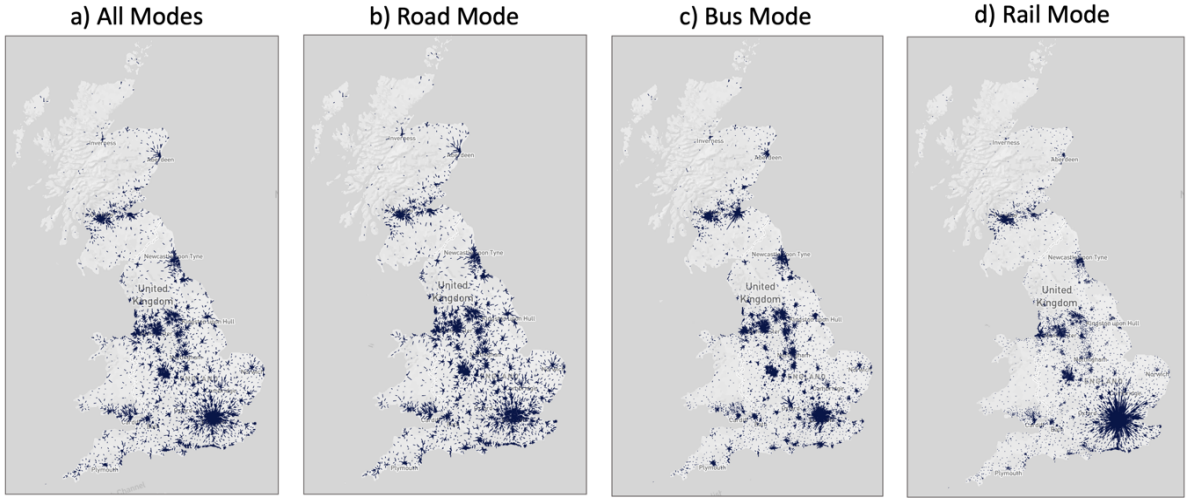

*Figure 3: Vector Flow Averages for All Modes, Road, Bus and Rail for GB*

We illustrate the three modal vector flow patterns based on these averages in Figure 3 where the differences in the underlying networks are very clearly reflected in these structures. When we aggregate all modes to $T_{ij} = \sum_k T_{ij}^k$ whose average flows are shown in Figure 3(a), these are quite similar to the flows on the road networks, that is $T_{ij}^{road}$ which we show in Figure 3(b). The flows on the bus network $T_{ij}^{bus}$ in Figure 3(c) are more compact thus illustrating that these are more concentrated in the cities, whereas the rail networks $T_{ij}^{rail}$ in Figure 3(d) show the dominance of rail travel in the big cities, particularly in the London region. To an extent, the relative proportions of trips by modes $\sum_{ij} T_{ij}^k / \sum_{ijk} T_{ij}^k$ are 79% road, 10% bus, and 11% rail. When we add the active travel layer to this, then, these proportions change to 69%, 12%, and 5% with active travel equal to 14%.

## Tuning the Twin to the Real System

The model-building process consists of a sequence of stages that tune or train the model to reproduce certain salient statistics of its observed distributions, and thus the model can be validated by comparing observations with the predictions based on the parameters that are derived from the process of tuning. The best process of validation is one where the fitted model is applied to an entirely different set of data from the one used to train it but in the past this has been almost impossible due to the fact that most urban models of this kind have been massive, often unwieldy projects built solely for individual cities without the resources to repeat the same model for different city applications. Data too for different cities has been problematic and only in the last decade have data sets become available which enable consistent applications across different cities. However we now have the computational and data resources to be able to build an original twin for England and Wales (Batty and Milton, 2021), then to extend this to Scotland where we have merged all three countries in a composite model. The basic tuning involves estimating the modal parameter values $\lambda^m$ for rail, road and bus that ensure that the average travel times $\overline{C^m}$ for these modes are reproduced. These values dimension the model and although they are intimately associated with the way these spatial interaction models are derived and fitted (using entropy maximising, for example, see Wilson, 1970), the tuned models are not necessarily the best fits to the data for the model performance can vary across different statistics. In this section, we will define these statistics and discuss their fit to data but note that the idea of testing the model against a different application is impossible when we work at the scale of a country. Comparable countries on which to test the model do not exist.

The residential location model with parameters $\lambda^m(\tau)$ begins at iteration $\tau = 1$, first generating a mean trip length or travel time/cost $\overline{C^m(\tau)}$ defined from equation (6) as

$$T_{ij}^m(\tau) = E_i \frac{A_j \exp(-\lambda^m(\tau) c_{ij}^m)}{\sum_z \sum_j A_j \exp(-\lambda^z(\tau) c_{ij}^z)} \quad \text{, and } P_i(\tau) = \sum_j T_{ij}^m(\tau) \quad , \qquad (11)$$

$$\overline{C^m(\tau)} = \frac{\sum_{ij} T_{ij}^m(\tau) c_{ij}^m}{\sum_{ij} T_{ij}^m(\tau)} \quad \text{where the observed mean is } \quad \overline{C^m} = \frac{\sum_{ij} T_{ij}^m c_{ij}^m}{\sum_{ij} T_{ij}^m} \quad . \qquad (12)$$

If $\overline{C^m(\tau)} > \overline{C^m}$, then with the parameter defined as a negative exponential, we need to increase these parameter values through the sequence implied by

$$\lambda^m(\tau + 1) = \lambda^m(\tau) \overline{C^m(\tau)} / \overline{C^m} \qquad . \qquad (13)$$

This iteration from $\tau$ to $\tau + 1$ will ultimately converge on the observed values $\overline{C^m}$, thus providing a baseline for the simulation.

The number of iterations needed to ensure that the parameter values $\lambda^m$ converge to within 1% of the observed values of average travel cost $\overline{C^m}$ is about 6 and as the current model runs within 40 seconds for all three modes, it takes just over 2 minutes to calibrate the complete model. As the model is designed to be run continuously so that users are able to test many different scenarios in interactive fashion, it only needs to be calibrated once. In designing 'what if?' type scenarios, the same parameters used to calibrate the model are usually used for each scenario, unless scenarios are explored which in themselves assume that the overall parameter values change. The calibrated parameter values are presented in Table 2 and these are entirely consistent with what we might expect from a system where the average travel times for road trips are some 53% less than bus trips which are 13% less than rail. Road trips by car average at some 12 minutes whereas by bus this is more than twice this average at about 26 minutes compared to rail at about 30. The parameter values reflect these average travel costs with much larger values for road than bus and then rail.

*Table 2: Calibrated Parameter Values and Mean Trip Travel Times*

|  | Road | Bus | Rail |
|---|---|---|---|
| *Observed Mean Trip Cost* $\overline{C^m}$ | 12.456 | 26.643 | 30.629 |
| *Predicted Mean Trip Cost* $\overline{C^m}$ | 12.553 | 26.676 | 31.004 |
| *Modal Parameter* $\lambda^m$ | 0.131 | 0.072 | 0.064 |

To compare the twin to its parent, there are many statistics comparing the goodness of fit between the real and simulated distributions and here we will use four different measures. The first two sets are based on variances and similarities which are usually computed for measures from aggregate data and here we will introduce these in generic form where we define each element in the observed distributions as $x_k$ and the predicted distributions as $y_k$. The first statistic is the standard correlation $\rho$ which is the ratio of the covariances between $x_k$ and $y_k$ and the individual variances, defined as

$$\rho = \frac{\sum_k (x_k - \bar{x})(y_k - \bar{y})}{\sqrt{\sum_k (x_k - \bar{x})^2 \sum_k (y_k - \bar{y})^2}} \quad , \quad -1 \leq \rho \leq 1 \quad . \tag{14}$$

$\bar{x}$ and $\bar{y}$ are the respective means of the observed and predicted distributions. If we square the correlation, we define the coefficient of determination $\rho^2$ which varies between 0 and 1, with a perfect fit at 1 and no fit at 0. Even though many distributions in the urban system are non-normal and highly skewed as we demonstrated earlier for the population and employment distributions across all 8436 zones of the GB urban system, these do not meet the standard statistical assumptions. However the correlation or rather its square is one of the most intuitively satisfactory statistics for measuring the goodness of fit as in its pure form, it accounts for the percentage of variation explained by the model in its logarithmic form.

The second statistic is also intuitively attractive for a perfect fit also accords to the value of 1 and no fit to 0. This is the Sorenson-Dice measure of similarity. In the variant used here (Masucci, et al., 2013), we examine each observation and prediction $x_k$ and $y_k$, first selecting the minimum of each, then taking the ratio of this minimum to the sum $x_k + y_k$ while scaling this by 2 to normalise this between 0 and 1. The statistic is thus

$$\vartheta = 2 \frac{\sum_k \min[x_k, y_k]}{\sum_k (x_k + y_k)} \quad , \quad 0 \leq \vartheta \leq 1 \quad . \tag{15}$$

This measure is not as crisp as the correlation for a large part of the calculation assumes that the model explains the minimum $\min[x_k, y_k]$ and its difference from a perfect fit is $\max[x_k, y_k] - \min[x_k, y_k]$. In the sorts of twin we are dealing with here, a large component of the explanation is bound to be accounted for, and this suggests that better measures might be normalised differences, although there is no consensus on this.

The second two statistics are quite widely used as measures of difference. The first of these is the root mean square deviation or error (RMSE) defined as

$$\phi = \sqrt{\sum_k \{(y_k - x_k)^2 / n\}} \quad , \quad \phi \geq 0 \quad . \tag{16}$$

Clearly a perfect simulation where predictions are the same as observations yields $\phi = 0$ but the range of the statistic is unbounded as this depends on the absolute values of the distributions. The fourth measure from probability theory is the information gain defined by Kullback (1959) as a divergence between a prior probability $p(x_k)$ – from the real observations – and a posterior $p(y_k)$ from the model's predictions. This can be seen as the difference or gain between the weighted value of the log of each observation and the weighted value of the log of each prediction which we state as $p(x_k) \log p(x_k) - p(x_k) \log p(y_k)$. We then define this as the expected value which is

$$I = \sum_k \{p(x_k) \log \frac{p(x_k)}{p(y_k)} \quad , \quad \sum_k p(x_k) = \sum_k p(y_k) = 1, \quad I \geq 0 \quad , \tag{17}$$

where the gain $I$ is unbounded.

There are two types of predicted distribution in the model – firstly, trip distributions by mode $T_{ij}^m$ of which there are three which we extend below to four, and aggregate trip distributions $T_{ij}$; then secondly, distributions of populations at destinations that we define as $P_j^m = \sum_i T_{ij}^m$ by mode and their aggregate equivalents $P_j = \sum_{im} T_{ij}^m$. Each of these distributions can be defined in terms of observations and predictions $x_k$ and $y_k$ and in the statistics defined and computed below, we substitute the actual observed distributions $x_k$, based on $T_{ij}^m, T_{ij}, P_j^m, P_j$ and their predicted equivalents $y_k$ into the four equations (14) to (17).

There are two further qualifications. First the predicted modal values do not sum to the observed because the model competes for trips for each mode, and although the total trips over all modes is the same as the observed, the total predicted over each mode does not in general sum to the observed modal total, that is

$$P^m(obs) = \sum_{ij} T_{ij}^m(obs) \neq \sum_{ij} T_{ij}^m(pred) = P^m(pred) \quad . \tag{18}$$

To correct for this and to ensure the predicted modal totals add to the observed, we need to multiply each predicted modal trip by the ratio $P^m(pred)/P^m(obs)$, that is

$$T_{ij}^m(pred) = \frac{P^m(pred)}{P^m(obs)} T_{ij}^m(obs) \quad . \tag{19}$$

This needs to be ensured for each mode so that any scaling differences are removed from the statistics defined above in equations (14) to (17). Second, in the digital twin, we are defining a very large number of potential interactions, over 71 million for each mode, and many of these values are very close to zero, below the minimum value that can be represented using logs in the machine, We therefore use the arbitrary rule that $if\ T_{ij}^m \geq 0\ then\ T_{ij}^m = T_{ij}^m + 1\ else\ T_{ij}^m = T_{ij}^m$. This is a quick fix, a not entirely satisfactory hack but it suffices to enable the statistics to be computed.

The key statistics are estimated and illustrated in Table 3 for the original data where the same set of statistics are defined using the logarithms of the same data. In estimating the performance of the model, it is clear there is no best or unique statistic. The correlation and the Sorenson-Dice statistics show the kinds of performance that are typical of these types of model which is very roughly that some 50 percent of the variance can be explained at the level of the different modal interactions. In this variant of the model, we will not explore the disaggregate distributions of predicted populations at trip destinations because of difficulties over scaling noted above in equations (18) and (19) but in future variants of the twin, we will introduce constraints on the model that ensure that the different modal interactions sum to the observed system totals in a manner that does not compromise our reproduction of the mean trip lengths. The statistics for RMSE and information gain are more problematic with respect to the specific modal splits although in aggregate form for all trips and for populations, these show a correspondence between observed and predicted distributions which are consistent with the correlations and the Sorenson-Dice indices.

*Table 3: Basic Goodness of Fit Statistics for Predicted Populations and Trips*

|  | Predicted and Observed Modal Trip Distributions | | | | Population |
|---|---|---|---|---|---|
|  | $\sum_m T_{ij}^m$ | $T_{ij}^{road}$ | $T_{ij}^{bus}$ | $T_{ij}^{rail}$ | $P_j = \sum_{im} T_{ij}^m$ |
| *Correlation-Squared Statistics* $\rho^2$ | 0.832 | 0.679 | 0.562 | 0.396 | 0.685 |
| *Log* $\rho^2$ | 0.789 | 0.797 | 0.647 | 0.426 | 0.722 |
| *Sorenson Dice Statistics* $\vartheta$ | 0.606 | 0.436 | 0.417 | 0.214 | 0.866 |
| *Log* $\vartheta$ | 0.560 | 0.567 | 0.433 | 0.200 | 0.918 |
| *Root Mean Square Error RMSE* $\phi$ | 3.067 | 3.667 | 0.897 | 1.572 | 844.224 |
| *Log* $\phi$ | 0.292 | 0.231 | 0.152 | 0.307 | 0.164 |
| *Kullback Entropy Statistics I* | 0.0294 | 0.418 | 0.649 | 0.942 | 0.026 |

## An Interactive Web-Based Framework for Running The Twin

The model has been developed for the last 5 years with funding from various UK agencies (see Acknowledgements) and during this time, it has been extended from England and Wales to include Scotland. Currently we are adding a fourth/fifth network mode called active travel (cycle/walk) to the framework. From its inception, the model has been web-based (although local desktop versions do exist) and it uses a simple interface that enables any professional (informed) user to employ the model to explore the data and to test and evaluate 'what-if?' types of scenario. During its development, dramatic improvements in the speed of running the three mode version have been made in terms of programming and hardware utilisation (using GPU server chips) and currently the version that is online takes around 0.44 minutes to run, calibrate and refresh with respect to its basic graphics. This is using a 2017 MacBook as client and a high fibre domestic broadband which is a good deal slower than the research environment used for model development but it does accord to what typical users might have for access. The interface is still not as fast and seamless as we would like although the speed of access is being rapidly improved as we acquire better hardware. The robust version that any reader of this paper can get access to is at http://quant.casa.ucl.ac.uk/ although the results reported below, are based on later versions – QUANT2 for scenarios and QUANT3 for more experimental work. The version that is run for experimental purposes takes a matter of seconds but this utilises the optimised local hardware. Technical details associated with the twin are summarised in an earlier paper by Batty and Milton (2019) and the computational details, software systems used and the way the web services determine the client-server interface are covered in various notes that are accessible through the model's Github repository (https://github.com/casa-ucl/QUANT-UDL-Scenarios).

To give some idea of the look and feel of the interface, we show a sequence of shots from the user's screen in Figure 4. We cannot illustrate the entire interface here but we will sketch the main components in this collage. It begins with a splash screen which then throws up a toolbar that lets the user explore the data from thumbnail maps and then enables the user to fine tune or calibrate the model. This is shown in Figure 4(a). The performance of the calibrated model – that is the correspondence between observations and predictions of the existing system – can then be explored and here we show the predicted population accessibility and the deviations between observed and predicted in Figure 4(b). The observed population accessibility is shown earlier in Figure 2(b) and the deviations in 4(b) are computed from the data in these two maps. As we noted above, there are many other ways of comparing model fit and the interface shown here enables the user to access visual and statistical information associated with this process. The model has many other layers at national level which are represented in MSOA form, and we show one of these – Green Belts (including Areas of Outstanding National Beauty) – in Figure 4(c) with the inset being the Oxford Green Belt which we use in the illustration of a development scenario below.

Before the model is calibrated through fine tuning its modal parameters to meet the observed mean trip lengths, there is also an obvious possibility for exploring the structure of the observed data visually and numerically. We have already indicated some of the analytics that pertain to the distribution of population, employment and trips in an earlier section, but this can be taken a lot further either within the interface (implied by the functions in Figure 4) or off-line by downloading the data and engaging in various kinds of spatial analytics that involve examining the relationships between different spatial aggregations and their patterns (Boeing, Batty, Jiang and Schweitzer, 2023). Once these early stages are complete and they only need to be engaged once, the user can move on to generating scenarios, with the rest of this paper illustrating typical "what if?" scenarios which can affect the entire country. Before we do this, it is worth sketching the range of scenarios that are possible with the model and also noting that each scenario can be thought of as a variant of the model, another twin if you like. This implies that in urban models of this kind, any variant can be regarded as a twin and that the process of exploring urban futures and the future of cities using simulation of this kind is always likely to be a demonstration of the fact that many different twins exist for the same problem.

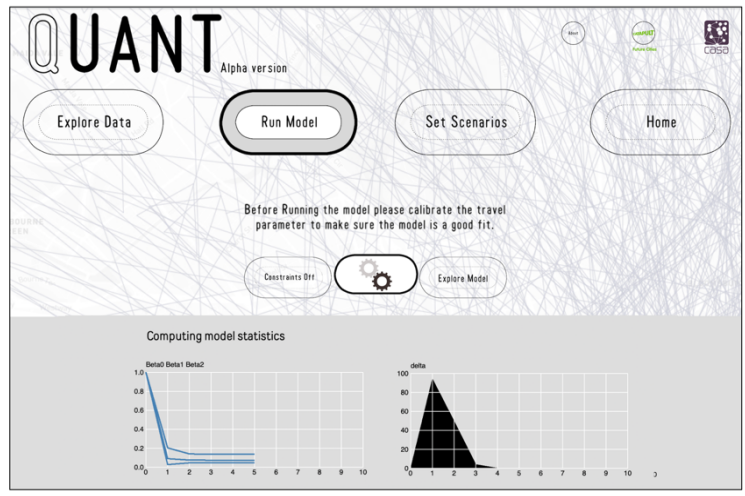

a) Tuning and Testing the Model

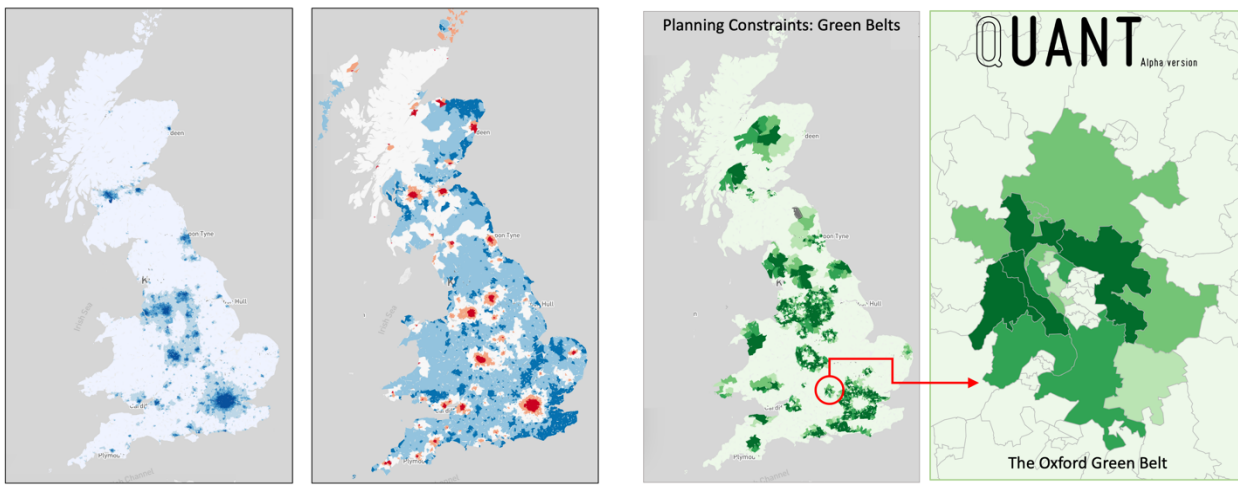

b) Observed & Calibrated Job Access $\quad$ c) National Green Belts

$$V'_i = \theta \sum_j P'_j c_{ij}^{-2}$$

Residuals $V'_i - V_i$

*Figure 4: Typical User Interaction with the Model, its Data and its Predictions*

When the model is calibrated with respect to its parameter values $\lambda^m$, we can begin to explore scenarios moving to the screens which we show in Figure 5. In this figure, we show the impact of locating three new sites each with 5000 new jobs west of Oxford which are partly constrained by the Oxford Green Belt. The predicted populations associated with these increases in jobs are located east and north of the job locations showing how the configuration of existing population influences these impacts. We do not have the space here to explore these implications in detail but the key idea is clear. This figure simply shows what the user using the model would see in successive screen shots and it implies that any scenario should be built by changes to the key independent variables that determine location and interaction as we implied above. Any of the model's inputs can become the basis for a new scenario such as this which can involve changes to employment at origins $E_i$, modal trip costs in terms of travel time $c_{ij}^m$ associated with the parameter values $\lambda^m$ which reflect the impact of travel times on mobility, and the imposition of constraints on population $P_j$.

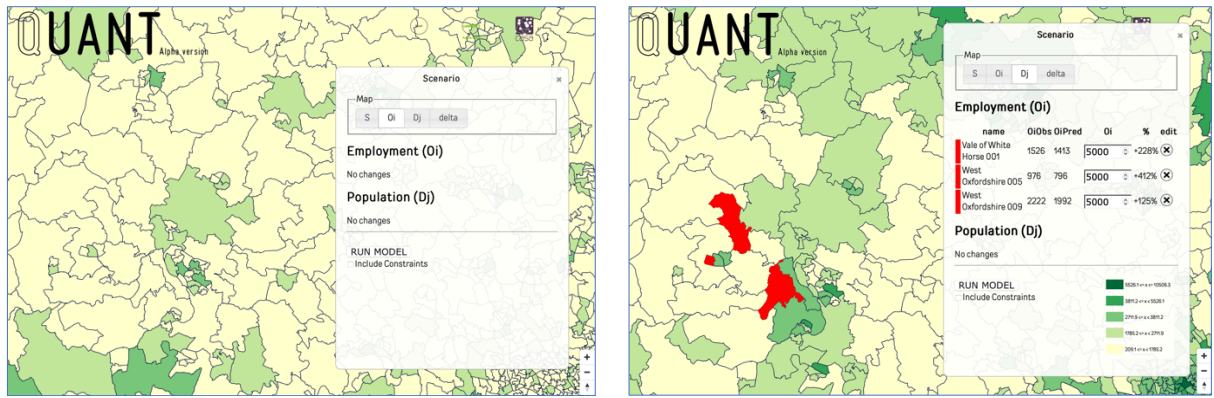

*Figure 5: Generating The Impacts of New Jobs in the Oxford Region*
*a) Left: Observed Population b) New Jobs (Red) and New Population Impacts*

There are an enormous number of changes to these values that form the elements of any scenario; for example there are $8436^2 = 71{,}166{,}096$ interaction elements that can be changed and manipulated to generate an almost uncountable infinity of possible scenarios. Add to this, changes in the amounts of the 8436 employment and populations volumes and the problem explodes in an entirely unmanageable way. In fact, all we have ever done with this model is test very simple scenarios with no more than 100 or so changes and even this is quite a large number for any scenario to embrace. If we examine any town or regional plan, it is likely to have a multitude of changes which are defined in a fairly fluid way but when these are then compared with each other, even scenarios with relatively few changes to the current baseline generate many impacts that need to be assessed comprehensively. Before we explore the use of the twin to generate and test urban futures for GB, we will illustrate how particular variants of the model – many twins that emerge when we alter the form of the model's spatial representation and interaction – provide applications of the same model, thus generalising the whole question of what a twin is and how many there might be based on the same functional structure of a particular model.

Generating Multiple Urban Twins

The most transparent way of generating variants of a digital twin of a system of cities is to apply it to different spatial representations. This means the twin is developed for different spatial scales which are usually defined using different numbers of subdivisions or zones reflecting different hierarchies of subsystem. In our application here, this involves defining different geographies which in turn are composed of different output areas which define the standardised geography of the Population Census but there are multiple other geographies. These range from dividing the system into regular tessellations such as those based on a grid all the way to abstracting the space as a collection of subsystems without specific dimension. These are equivalent to the nodes in a graph whose linkages represent the topology rather than the planar landscape which we usually employ in the applications to cities and regions like those developed here.

In essence the model remains the same in terms of its function, structure and operation, in terms of its algebra, and estimation but the system to which it is applied, changes in form. In the sense that form follows function, the predictions of the model are likely to change too with the general caveat that models with ever more zones imply ever more detail and are harder to fit than those with lesser numbers of zones. When the number of zones is aggregated to one, the functional structure of the model collapses to a single location and the model usually achieves a perfect fit. This of course is a kind of "argumentum ad absurdum" which makes the model meaningless.

We first illustrate this notion of changing the model's form by breaking it up into different subsystems that are adjacent but separate from one another. The database at the core of the GB model consists of zones (MSOAs) which comprise the three countries that make up GB – England, Scotland and Wales. The total number of zones for GB is 8436 as noted but these can be broken down into 6,791 for England, 1235 for Scotland and 410 for Wales. We can run for the model first for the individual countries and from the earlier analysis of density and size, we are able to speculate that that the travel patterns in England are more compact and the densities of employment higher than the same variables defined for Wales which in turn is greater than those for Scotland. We would expect this to be reflected in the outputs – the predictions of residential population and the modal interactions which we will measure somewhat coarsely in this section which is a first foray into differences posed by changing the model's spatial extent. We can also combine England and Wales into a single model with 7201 MSOAs and England and Scotland into another with 8026 MSOAs but we cannot link Wales to Scotland as these are disconnected in terms of a lack of contiguity between MSOAs in Wales with those in Scotland. In fact in models of this type, we are able to have disconnected zones in theory and in certain applications, it might be necessary to simulate zones that are disconnected from one another; but in all the examples here, this is not the case.

Key statistics based on how well each variant performs relative to its data are shown in Table 4 where we show the mean trip lengths for the three modes of each model $\overline{C^m}$, the related parameter values $\lambda^m$, and the correlation $R^2$ between observed and predicted residential population $P_j$. The differences are not trivial but whether or not they imply that the models of each variant are sufficiently different from one another and from the original model for GB not to be defined as 'twins' remains an open question and one that appears largely unanswerable. These differences are buried within the data when aggregated or disaggregated to different scales and it is not possible to generalise as to how these determine differences between the models. It appears that as the system gets larger from Wales to Scotland, to England, to Wales and England, to Scotland and England and thence to GB, the travel patterns get more dispersed for road and more concentrated for rail. Of course the relative performances of the models improve the smaller they are but the picture is not easy to disentangle. Surprisingly, there has been hardly any work done on generating variants or twins of many different types of comprehensive or partial urban models, such as those based on discrete choice theory, agent-based simulation and cellular automata, and thus the idea of multiple digital twins at this point can only rest on the model types introduced in this paper. As we aggregate the data to generate families of twins, their massive variability which is hidden when data is aggregated, requires much more complex data analytics for their unravelling and this is a research direction that has also rarely been considered prior to the idea that there may be many rather than a single model or digital twin.

Our second approach to defining different variants of the basic GB model involves moving to a different level of spatial disaggregation. We are not able to demonstrate the model at the level of the lower super-output area (LSOA) which would mean moving from 8436 to 41,729 zones and 1.74 billion potential interactions, This LSOA model is under development but currently it is not available to our online users. Therefore to demonstrate this at the lower level, we are only able to do this for Greater London and thus we have built the model first for 983 MSOAs that define the metropolitan GLA area and then for the 4835 LSOAs that involve a detailed nested disaggregation of the MSOA level. It is this disaggregation that changes the nature of the data much more substantially than partitioning the area up into subsystems such as cities or countries. The first comparison is between the GLA and the baseline GB model where from Table 4, we see that the average mean travel time on the road for car users drops dramatically some 50% to about 6 at the GLA level from 12 minutes. This is primarily accounted for by the much greater density of population and smaller scales of zones at the urban rather than at the national level, and this is also reflected, although not as dramatically, for bus travel and rail travel which also fall between 10% and 45%. This is sufficient to show that moving from the national to local urban scale involves a big change in the way populations travel with much more flexibility in road transport than bus or rail which depend on fixed infrastructures.

*Table 4: Mean Trip Cost $\overline{C^m}$, Population $R^2$, Calibrated Parameter Values $\lambda^m$ (in brackets)*

| Spatial Representation of the System | Road | Bus | Rail | $R^2$ Population |
|---|---|---|---|---|
| GB: England, Scotland and Wales | 12.457 (0.131) | 26.643 (0.072) | 30.628 (0.072) | 0.685 |
| England | 11.936 (0.139) | 26.292 (0.075) | 28.575 (0.072) | 0.606 |
| Scotland | 12.023 (0.117) | 21.599 (0.080) | 29.311 (0.052) | 0.714 |
| Wales | 10.969 (0.139) | 19.799 (0.082) | 28.847 (0.047) | 0.699 |
| England and Wales | 12.168 (0.136) | 26.477 (0.074) | 30.066 (0.066) | 0.617 |
| England and Scotland | 12.237 (0.133) | 26.458 (0.073) | 30.374 (0.065) | 0.684 |
| Greater London (MSOA) | 5.947 (0.295) | 25.135 (0.082) | 21.582 (0.057) | 0.578 |
| Greater London (LSOA) | 5.913 (0.294) | 23.050 (0.071) | 15.218 (0.078) | 0.556 |
| GB with Active Travel (and Calibrated $\lambda^m$) | 12.457 (0.136) | 26.643 (0.073) | 30.629 (0.065) | 0.676 |

When we move between output area geographies, there are much more substantial changes in terms of the geometry of spatial representation. In terms of the GLA, moving from MSOA to LSOA involves increasing the number of zones (polygons) from some 983 to 4835 which is nearly a 5 times increase. The degree of detail picked up from the different modal networks occasioned by this change is considerable with the average area of zones changing from 159.934 to 32.516 hectares and average population from 8315 to 1691 both reflecting this 5 times multiplier. We show the two spatial systems in terms of their zones in Figures 6(a) and 6(b) although we are not able to show the density of the transport networks at the level of resolution of the page. In fact the biggest changes in the way the data and spatial representations are assembled involves the way the shortest routes between zone centroids are computed. The centroids reflect the detail of urban development in each polygon and the way these are weighted relative to the detailed networks that tie these developments together. We have not explained how the centroids or the shortest routes between them are built from the very detailed modal networks and this involves a number of stages that involve linking rail, bus and road network nodes and segments from the most detailed scale at which these networks are available. This involves setting various thresholds to determine if population is able to access bus stops and stations through access to the road network which is the base layer, and of course this changes substantially between the MSOA and LSOA representations of Greater London.

To an extent, the performance of the MSOA and LSOA variants is reflected in the areal and population/employment changes between the two levels. From Table 4, it is clear that the change in access to modes between the two levels in terms of the mean travel is quite small for the road and rail modes. It is only about 1% for road, more for bus at about 9% but the biggest change is for rail which declines in mean trip length by 30% . We hazard a guess that this is due to the fact that in central London, the rail network is very highly concentrated, more so most other big cities, with many more users than in the inner and outer suburbs, and these are clearly reflected in the model's parameters. In terms of the MSOA model, there is only one zone unreachable by road and one by bus but 571 out of 983 unreachable by rail that account for the sparsity of the rail network and its stations. This problem becomes critical when we move to LSOA level, in that 4 out 4835 zones are unreachable by road, 604 by bus but a massive 4316 out of 4835 unreachable by rail which is almost 90 percent. This suggests that in developing this model further we must move to a much more detailed mixed mode set of models where population can travel as far as they like by accessing appropriate and connected modes.

In fact to make more detailed comparisons of model performance, we plot the predicted populations for each of the models in Figure 6(c) and 6(d) where we aggregate the population data for the LSOA model to MSOAs in Figure 6(c). This gives us a direct visual comparison between 6(c) and 6(b). The correlation $R^2$ between these maps layers is 0.937 which is one of many measures of similarity between the two variants. Whether or not we can say anything from this about the models being 'the same' rather than 'similar' is back to the open question that we raised earlier: whether or not every variant of a basic model is a version or twin of the same or a different model? In the next and final substantive analysis of the twin, we will explore the idea that a scenario can be a twin in that typical scenarios with models of this kind, manipulate the spatial representation in ways that involve changes to travel times, activity locations and constraints on development. Manipulating changes in the functions defining the model might also be regarded as more fundamental model scenarios but we will not extend our model to deal with these kinds of future.

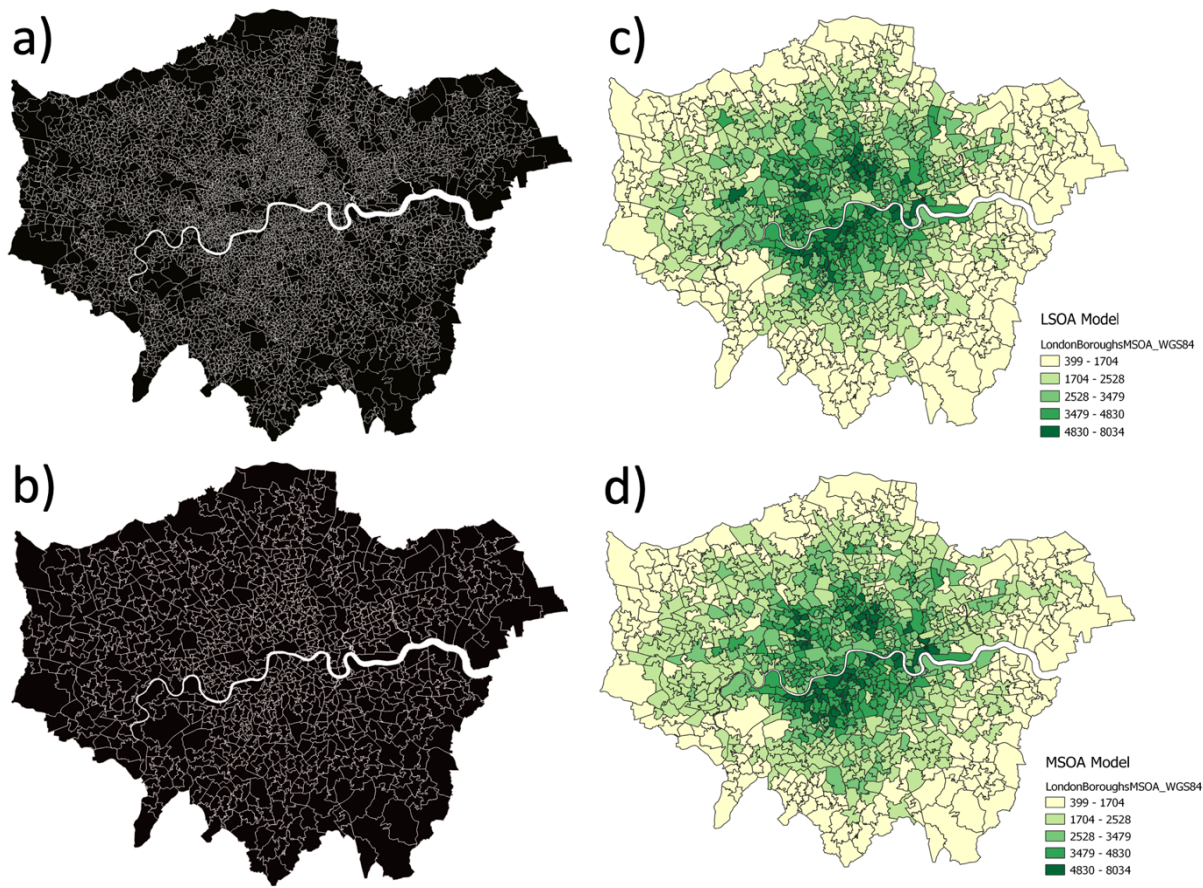

Figure 6: Different Levels of Spatial Resolution Lead to
Different Model Performance
a) LSOAs b) MSOAs c) Population Predicted at LSOA Summed to MSOA d) Population at MSOA

*Scenarios as Twins: Exploring Transport and Urban Futures for GB*

Almost 200 years ago in 1825, the first passenger railway was opened between Stockton and Darlington in Northeast England and this ushered in a half century of railway building when the entire country including the island of Ireland was dominated by the construction of new lines. In fact, in the 1840s, this was referred to as 'Railway Mania'; for example, in 1846, more than 250 bills were proposed and passed by Parliament for new tracks (Wikipedia, 2023b). By the 1880s however, construction had virtually ceased and since then hardly any new lines have opened. In fact during the 1960s under the Beeching Plan (British Railways Board, 1963), many were closed and only since the advent of high speed trains in the last 20 or so years, have there been any new proposals to regenerate and modify the network. A wave of new construction has been recently proposed under a comprehensive but highly controversial Integrated Rail Plan (DfT, 2022) from which a small amount of work has already been initiated.

The plan is built largely around a new line referred to as HS2 (High Speed 2 where HS1 is the Channel Tunnel line). This line was originally proposed to run from London to Birmingham and thence to Manchester with a spur to Nottingham as shown in Figure 7. It was supported by upgrades to the east coast mainline from London to Newcastle and on to Edinburgh and a similar upgrading of the west coast line from Manchester to Glasgow. The East Midlands line east of HS2 from London to Sheffield was also marked for upgrading and the upgraded Liverpool to Leeds line – the so called Northern Powerhouse Rail – more or less completes the project. We show these lines and improvements in travel times in Figure 7(a) where our first test in Figures 7(b) and (c) will be based on this complete plan. In fact as some readers (specifically in the UK) will know the plan has been dramatically scaled back with the Nottingham and Manchester lines now abandoned and the only new line being between London and Birmingham. Even that line is widely regarded as not linking the most appropriate locations together. What we will do here is test the complete plan, then pare it back to simply the London-Birmingham leg, and then measure the impacts of this plan. To do this, we will generate changes to the mean trip travel times but more particularly examine the modal shifts and carbon saved from these lines which we will detail as follows.

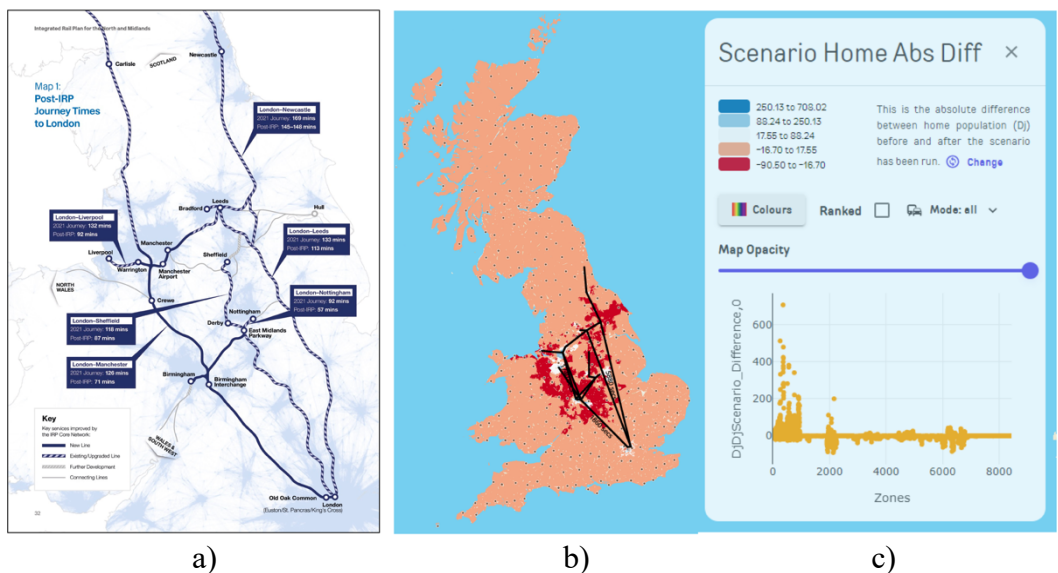

a)  b)  c)

*Figure 7: The Integrated Rail Plan and Its Spatial Impact on Populations*

The Plan Published 2021 by UK Department for Transport b) Displacement of Populations With the Plan Fully Implemented, and c) Levels of Displacement (Population Loss > -16 (red); Population Loss Gain >-17-88 (white); Gain >88-250 (light blue); Gain>250 (blue))

There are many measures of impact related to how these new rail lines reduce travel times and costs, thereby displacing population by shifts between different transport modes, changing the distances in kilometres travelled on different modes. We can also compare these changes with the costs of implementing different elements of the plan and in this way calculate cost-benefit (CB) statistics for different modes. We will not compute these statistics here, largely because the costs are so indeterminate and the construction costs of new and renewed rail are only a small part of the total that are difficult to estimate. Here our two examples are as simple as possible, composed of plugging in the IRP and then HS2 Phase 1 as two scenarios to be tested by the model, thus producing population shifts from changes in all three modes and working out the number of kilometres saved (or increased where travellers are forced onto less efficient modes). As one might expect, the IRP has a much bigger impact than the HS2 Phase 1 as there are many more rail lines and this is clear from comparing Figure 8(a) and 8(b). These maps show that the hubs in both cases where new lines terminate or cross, attract population while in the wider hinterlands around the lines, there is substantial loss of population. The hubs in the IRP all attract population and this implies that the new lines lead to urban concentrations, arguably adding to economies of scale in the largest cities. In the case of HS2, the biggest impacts are in the London hub where population is attracted from surrounding areas while in Birmingham and the Midlands there is substantial loss of population.

In the case of the IRP, almost 1% of the total trips (19.952m in GB) are attracted to the new rail lines while some 0.07% and 0.2% of trips are lost from road and bus respectively. This leads to an increased daily travel of 35.446 million kms on rail and a reduction of some 2.355m and 0.563m kms on road and bus. The total shift involved here is 3.266 greater than the HS2 Phase 1 scenario where some 0.02% of the total trips are attracted to the new rail while some 0.01% and 0.005% of trips are lost from road and bus respectively. This is much less than the IRP but to take the analysis further, we would need to explore the very detailed locational and trip volume shifts at the level of MSOAs. We cannot do this here for all we have time to show is the kind of analysis that can be generated by the model. In our current work, we are improving the impact statistics and adding full costs of travel as well as construction to the analysis.

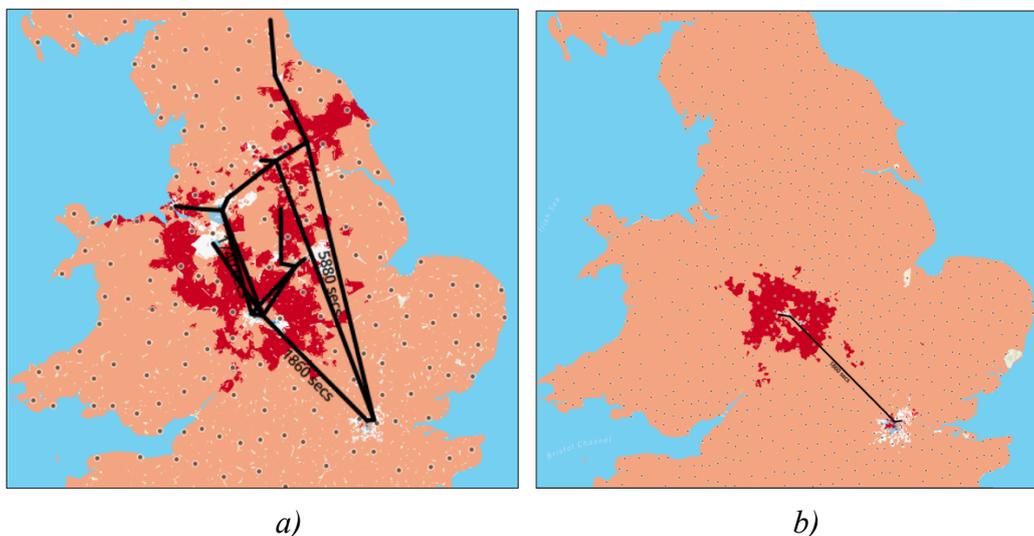

*a)* *b)*
*Figure 8: Zooming in on the Displacement of Population a) The Integrated Rail Plan, b) HS2 First Phase*

So far we have not developed a detailed and comprehensive set of indicators which can be used to compare different scenarios. In short, we cannot say one is any better than any other and to do this, we need to transform the measures that we are developing for each scenario into measures of optimality. This is a major goal for future development of the framework and it will require us to ground the model in different measures of cost-benefit and consumer surplus (Glaister, 1981). We are also working on generating literally thousands of alternative scenarios, using various frameworks which evolve new scenarios from those generated so far which improve their optimality using various artificial intelligences that build in successive improvements akin to methods in machine learning for propagating solutions that build on architectures such as neural nets

## Conclusions and Next Steps

Physical or spatial scale is a key organizing principle for models that range from buildings to entire cities and thence to nation states, continental regions, and to the earth itself. There are now a plethora of digital twins that are being developed for the earth system such as those developed by NASA (2023) and the ESA (2023), and for city states such as Singapore (Davison, 2022; MPA, 2023). In cities and related spatial systems, changing the scale does not in essence change the nature of the model. As we have shown, when we aggregate from the finest scales of building to entire cities and thence to regions and nations, there is nothing in a model that need change other than its representation. In this sense, we can have many digital twins of a real system at different scales without there being any difference in the system's actual functionality at these different scales. In fact due to the somewhat ambiguous meaning of the term digital twin, subtle changes can occur as scale changes but even for the narrowest of real systems which are entirely physical in form and function, each variant of the digital twin can manifest differences that could be regarded as generating a different digital twin.

Scaling a model from the building level to the city and upwards is a key issue in the model developed here which has been designed to directly address the fact that the real city systems and its environments are often impossible to bound. It is hard to provide clean breaks between a city and its environment and in a global world where many activities are now widely spread spatially, there is an unassailable logic in finding physical boundaries that constrain interactions. An island nation like Great Britain is uniquely configured to contain most of the spatial interactions between its 20 million workers within its physical bounds. There is still the problem of defining cities within this area but as we have shown, our computing resources have grown to the point where we can run rather large models with numbers of zones in the order of 1000 and interactions in the order of 100 million quickly and efficiently. At this level, we can replicate very large systems at a relatively fine scale.

What we have not hinted at here but need to do so with respect to continued development of the QUANT framework involves coupling models together. Not only might there be many different models of the same real system, there is now strong momentum as computation gets ever more powerful to put models together; to integrate different functions within different models or to simply string models together. This can yield chains of digital twins, perhaps even arranged as hierarchies. In this sense, any problem context which generates the need for formal models opens up the possibility of different kinds of model integration and coupling and if we add differences in scale to this kind of elaboration, we quickly merge these ideas into those that involve the creation of integrated platforms where a variety of digital twins can be fashioned into powerful federations of model. These possibilities are beginning to guide the way what in the past were individual models can be generalised to much bigger spatial systems as well as their spreading out into wider computational environments.